# Orbital angular momentum accumulation in $SrVO_3$ thin films

Julien Brehin[1], Montserrat X. Aguilar-Pujol[1], Dongwook Go[2,3], F. Casanova[1], J. Fontcuberta[4], E. Longo[4*]

*1. CIC nanoGUNE BRTA, E-20018 Donostia-San Sebastian, Spain*

*2. Department of Physics, Korea University, Seoul 02841, Republic of Korea*

*3. Center for Quantum Dynamics of Angular Momentum, Pohang 37673, Korea*

*4. Institut de Ciencia de Materials de Barcelona (ICMAB-CSIC), Campus de la UAB, Bellaterra (Barcelona), Spain*

Corresponding author: * elongo@icmab.es

## Abstract

Orbital transport in light transition metals has emerged as a promising route toward angular-momentum electronics, with Hanle magnetoresistance (HMR) providing a direct electrical probe of the orbital Hall effect (OHE) in non-magnetic conductors. Here we report magnetoresistance signatures consistent with orbital HMR in epitaxial $SrVO_3$ (SVO), a narrow-band $d^1$ oxide grown on (001)-oriented $(LaAlO_3)_{0.3}(Sr_2TaAlO_6)_{0.7}$ substrates. In films of different thickness and longitudinal resistivity, field-dependent measurements reveal a reproducible positive, even-in-field and approximately quadratic $\Delta\rho_{x-y}$ response, as expected for HMR. Angular measurements characterize the corresponding magnetoresistance anisotropy and reveal an additional non-Hanle contribution. Density functional theory (DFT) calculations yield an intrinsic orbital Hall conductivity $\sigma_{OH}^{DFT}$ = 390 (ħ/e) $\Omega^{-1}cm^{-1}$ and a spin Hall conductivity $\sigma_{SH}^{DFT}$ = −12 (ħ/e) $\Omega^{-1}cm^{-1}$ at the Fermi level, corresponding to $|\sigma_{OH}^{DFT}/\sigma_{SH}^{DFT}| \approx 33$, thus indicating a predominantly orbital response. Using the diffusive HMR framework with $\lambda_{OD}$ = 2 nm as a reference value, the 20.8 nm film gives conservative saturation-limit lower-bound (LB) estimates $\theta_{OH,LB} = 0.0144 \pm 0.0004$ and $\sigma_{OH,LB} = (460 \pm 11)$ (ħ/e) $\Omega^{-1}cm^{-1}$, comparable to the DFT value. Across the series, lower-bound estimates obtained with the same $\lambda_{OD}$ increase overall with longitudinal conductivity. These results suggest that narrow-band $d^1$ metallic oxides are a promising platform for orbital transport.

## 1. Introduction

The orbital Hall effect (OHE), the orbital analog of the spin Hall effect (SHE), has emerged in recent years as a fundamental contributor to angular momentum transport in solids.[1–3] Unlike its spin counterpart, the OHE does not require strong spin-orbit coupling and is predicted to produce orbital Hall conductivities $\sigma_{OH}$ that exceed the spin Hall conductivity $\sigma_{SH}$ in light 3d transition metals by up to two orders of magnitude. This makes 3d metals attractive candidates for orbital current generation and detection.

Experimental evidence for the OHE in 3d metals has recently been reported using several complementary techniques: magneto-optical Kerr effect in Ti, Cr and V;[4,5] spin-torque ferromagnetic resonance in Ti and V;[6,7] THz emission in 3d and 5d metals;[89] and most relevantly for the present work, Hanle magnetoresistance (HMR) in Mn[10,11] and in V.[12] HMR is a particularly useful diagnostic, because it can allow simultaneous quantification of the orbital Hall angle ($\theta_{OH}$), the orbital diffusion length ($\lambda_{OD}$), and the orbital Hall conductivity $\sigma_{OH}$ from a single transport measurement in a single layer material. Because no ferromagnetic layer is involved, HMR is also less susceptible to interface-related artefacts.

Recent experiments indicate that orbital transport and relaxation are strongly material and disorder dependent. In Mn, orbital-HMR measurements across films with different crystalline order show an approximately linear scaling of $\sigma_{OH}$ with $\sigma_{xx}$ in the disorder-dominated regime, together with substantial changes in the orbital relaxation time.[11] In $SrRuO_3$, orbital-torque measurements instead reveal an unconventional behaviour in which $\sigma_{OH}$ is nearly constant at high conductivity and increases as $\sigma_{xx}$ decreases, attributed to a Dyakonov–Perel-like orbital-relaxation mechanism.[13] These contrasting behaviors suggest that orbital relaxation cannot be described by disorder alone, but may also depend on the underlying electronic structure. In this respect, Mn, $SrRuO_3$ and $SrVO_3$ (SVO) span markedly different orbital and bandwidth regimes, with SVO characterized by a narrow $t_{2g}$ manifold.[14,15] This motivates examining whether electronic bandwidth, orbital hybridization and correlations contribute to the markedly different orbital-relaxation and conductivity-scaling behaviors observed across these systems.

A striking feature of several OHE measurements in 3d metals is the substantial discrepancy between the experimentally extracted $\sigma_{OH}$ and first-principles predictions. In vanadium, $\sigma_{OH}$ of the order of 4500 - 6050 $(\hbar/e)\ \Omega^{-1}\ cm^{-1}$ is theoretically predicted, whereas the value reported from HMR by Aguilar-Pujol et al. is 76 $(\hbar/2e)\ \Omega^{-1}cm^{-1}$, i.e. 38 $(\hbar/e)\ \Omega^{-1}\ cm^{-1}$,[12] a suppression of more than two orders of magnitude. This discrepancy has been attributed to disorder-induced vertex corrections, which account for the modification of the intrinsic Hall response by repeated impurity scattering,[16] and to quantum-kinetic treatments of orbital transport, which distinguish the orbital response relevant to nonequilibrium transport and accumulation from the conventional intrinsic Kubo orbital Hall conductivity,[17] both of which suppress $\sigma_{OH}$ in light centrosymmetric metals.

A central question for the field is therefore whether the OHE detection and quantification *via* HMR can be extended to materials systems where the orbital character is qualitatively different from that of pure metals, and whether disorder-induced suppression is universal or system-dependent. Narrow-band $d^1$ transition-metal oxides offer a natural platform to address this question. The perovskite SVO is a moderately correlated metal with one electron occupying a narrow $t_{2g}$-derived manifold, close to a metal-insulator transition in the thin-film limit.[14,18–20]

Here we report magnetoresistance signatures consistent with orbital HMR in epitaxial SVO films grown on (001)-oriented $(LaAlO_3)_{0.3}(Sr_2TaAlO_6)_{0.7}$ (LSAT) substrates, with thicknesses from 6.5 to 20.8 nm. Field- and angle-dependent measurements reveal an HMR-like response in all films, accompanied by an additional non-Hanle contribution. Combining a diffusive HMR analysis of a 20.8 nm film with density functional theory (DFT) calculations of the orbital and spin Hall conductivities, we obtain conservative lower-bound estimates of the orbital Hall response comparable to the intrinsic DFT value, together with a sub-picosecond orbital relaxation time. Across the series, the extracted response correlates with longitudinal conductivity, a trend we treat as empirical because orbital relaxation may vary between samples.

## 2. Materials and Methods

Epitaxial SVO thin films with thicknesses ranging from 6.5 to 20.8 nm were grown on (001)-oriented LSAT single-crystal substrates by pulsed laser deposition at 750 °C under a controlled Ar atmosphere at a pressure of 0.03 mbar. Film thicknesses were determined by X-ray reflectivity for all samples except the nominal 10 nm film, for which $t = 8.2$ nm was estimated from the growth-rate calibration based on the XRR-measured films (Supplementary Table S1). The out-of-plane and in-plane crystallographic orientations of the deposited SVO films were determined by X-ray diffraction using θ-2θ scans and φ-scans, respectively (see Figure 1). The structural characterization of the complete film series is summarized in Supplementary Fig. S1 and Table S1. Hall bars with a length of 100 μm and a width of 20 μm were patterned by optical lithography followed by Ar+ ion milling. The insulating character of the LSAT substrate was verified after ion milling, ruling out possible parasitic conduction introduced during device fabrication. Magnetotransport measurements were performed in a Physical Property Measurement System (Quantum Design) over the temperature range 2-300 K and in magnetic fields up to ±9 T. Longitudinal and transverse voltages were measured using a DC current of 10 μA with a current-reversal scheme. For the 20.8 nm device, $R \approx 37.5\ \Omega$ at 100 K, corresponding to a Joule power $P = I^2R \approx 3.8$ nW. The low dissipated power, together with current reversal and repeated field sweeps, minimizes heating, voltage-offset and drift artefacts. Resolving a relative change $\Delta\rho/\rho = 3 \times 10^{-5}$ at this bias requires a voltage sensitivity of ≈ 11 nV. The longitudinal and Hall transport properties across the complete film series are summarized in Supplementary Fig. S2. A survey X-ray photoelectron spectroscopy (XPS) spectrum was acquired on an air-exposed LSAT/SVO(20.8 nm) film to check the surface chemical composition and detectable contamination by common ferromagnetic elements, as reported in Supplementary Fig. S3. Angle-dependent magnetoresistance measurements were acquired by rotating the magnetic field in three orthogonal planes: α (xy, in the sample plane), β (yz, with the current perpendicular to the rotation plane), and γ (xz, with the current parallel to the rotation plane), defined with respect to the Hall-bar current direction (x axis), as illustrated in Figure 2. The orbital angular-momentum polarization generated by the OHE is denoted by L and is oriented along the y direction.

## 3. Results

### *Structure and bulk transport*

Figure 1a shows the Bragg-Brentano θ-2θ scan of LSAT/SVO(20.8 nm) around the (002) reflections. The SVO film contribution appears as a shoulder of the LSAT substrate peak, close to the bulk SVO reference position at $2\theta \approx 47.28°$ ($c_{bulk}$ = 3.842 Å). For the 20.8 nm film, the extracted out-of-plane lattice parameter is c = 3.850 Å, corresponding to an out-of-plane nominal strain $\varepsilon_{zz} = (c - c_{bulk})/c_{bulk} = +0.22\%$. Across the films for which the SVO(002) feature can be reliably resolved, $\varepsilon_{zz}$ ranges from −0.20% to +0.22%, indicating only a minor out-of-plane lattice distortion and a nearly bulk-like c-axis throughout the series. The complete structural analysis is reported in Supplementary Fig. S1 and Table S1. The φ-scan of the (220) reflection (Fig. 1b) reveals four equally spaced peaks at φ = 1.7°, 91.7°, 181.6°, and 271.6° within

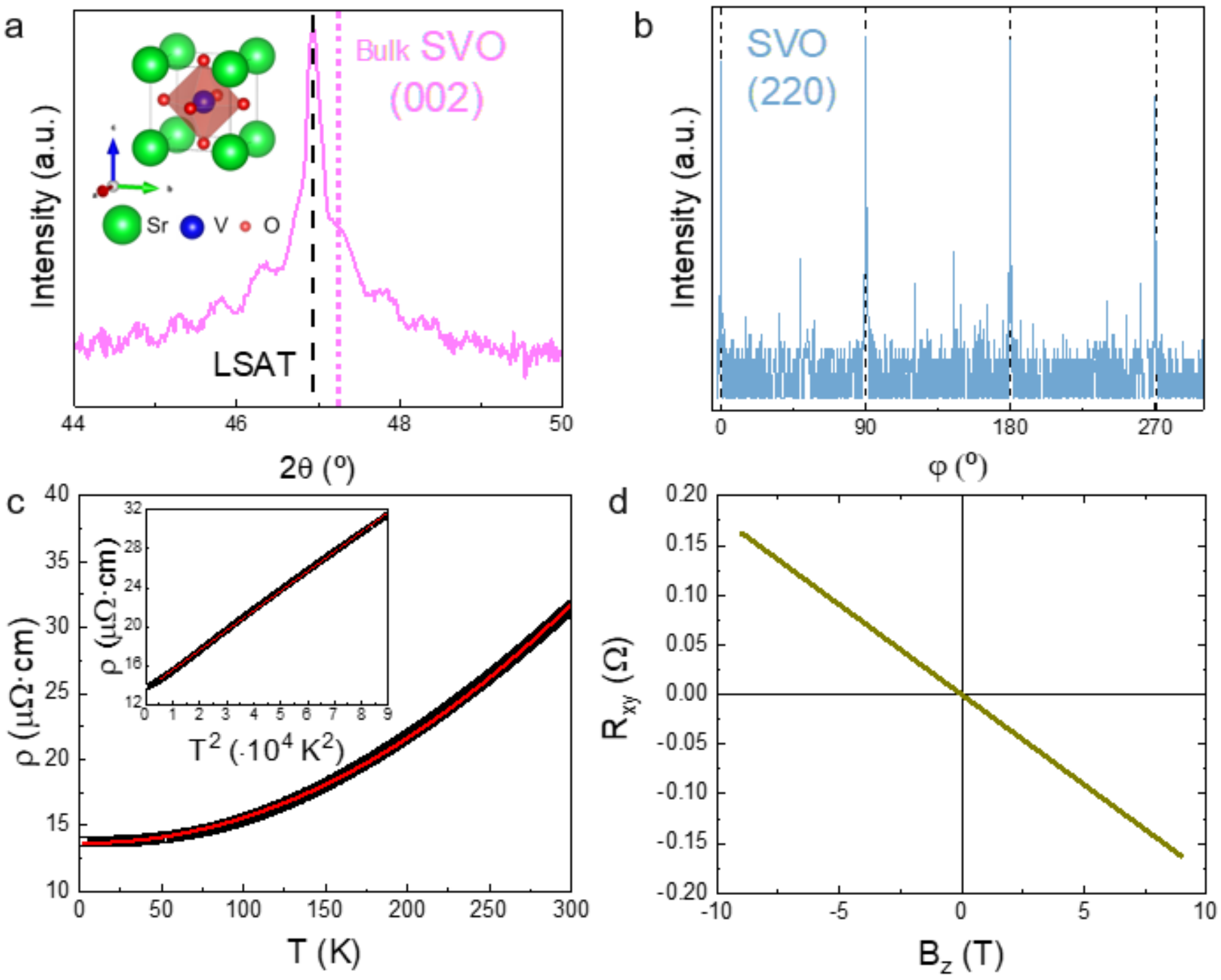

**Fig. 1. Structural and transport characterization of the LSAT/SVO(20.8 nm) thin film.** (a) Bragg-Brentano θ-2θ scan around the (002) reflections. Inset: cubic perovskite crystal structure. (b) Phi-scan of the (220) reflection. (c) Longitudinal resistivity ρ plotted versus T. The quadratic dependence is commonly observed in SVO films, indicated by the linear trend with R-square = 0.9995 when ρ(T) plotted versus $T^2$ (see inset). (d) Transverse Hall resistance $R_{xy}$ versus out-of-plane field $B_z$ at T = 300 K, showing a linear electron-like Hall response.

±0.1°, confirming the four-fold in-plane epitaxial symmetry. The patterned Hall bars exhibit metallic behavior with ρ(300 K) ≈ 32 μΩ·cm and ρ(2 K) ≈ 13.6 μΩ·cm, giving a residual resistivity ratio ρ(300 K)/ρ(2 K) = 2.3, comparable to literature reports of high-quality SVO thin films.[18,21,22] The ρ(T) dependence can be described either within a polaronic model[18] or by using a Fermi-liquid related form $\rho_0 + aT^2$, [15,23] over the entire measured temperature range. Within the simplest T-quadratic ρ(T) description, we obtain $\rho_0$

= 13.59 μΩ·cm and a = 2.02 × $10^{-4}$ μΩ·cm/$K^2$ ($R^2$ = 0.9996) (Fig. 1c). No resistivity upturn is observed down to 2 K, indicating that weak-localization (WL) and electron-electron interaction corrections are negligible in this film. The Hall resistance $R_{xy}$ is linear in $B_z$ up to 9 T at 300 K with a negative slope (Fig. 1d), indicating electron-like transport and no detectable anomalous Hall contribution. Consistently, survey XPS on an air-exposed film (Supplementary Fig. S3) shows only the Sr, V and O core levels and adventitious carbon, with no Fe, Co or Ni 2p signal above the background. A detailed temperature-dependent single-band analysis of the 20.8 nm device, reported in Supplementary Fig. S4, gives an effective Hall carrier density of about 1.2 × $10^{22}$ $cm^{-3}$ with only a weak temperature dependence. Representative Hall measurements across the broader series are shown in Supplementary Fig. S2. Because several $t_{2g}$-derived bands can contribute to transport, the single-band Hall density is treated as an effective transport parameter rather than as a stoichiometric carrier count.

### *Angular dependence of the magnetoresistance*

Figure 2 shows the angular dependence of the longitudinal magnetoresistance $\Delta\rho_L(\eta)/\rho = [R(\eta) - R(\eta = 90°)]/R(\eta = 90°)$, with η = α, β, γ, at T = 100 K and B = 9 T. The relatively high measurement temperature was selected to minimize potential low-temperature localization corrections. In both the α plane (xy, field rotated from the current direction (**J**) in the film plane) and the β plane (yz, field rotated from the out-of-plane direction within the plane perpendicular to current), the data exhibit the characteristic ∝ $\cos^2$ modulation expected for Hanle magnetoresistance.[24] We denote the fitted amplitudes in the three rotation planes as $A_\alpha$, $A_\beta$ and $A_\gamma$. The α-plane amplitude is $A_\alpha = (4.0 \pm 0.2) \times 10^{-5}$, while $A_\beta \approx 7 \times 10^{-5}$. The HMR scenario predicts identical amplitudes in the α and β planes and zero modulation in the γ plane (rotation in xz, where the field is always perpendicular to the orbital polarization **L** ∥ y). Our data depart from both expectations: the α- and β-plane amplitudes are unequal, and the γ-plane rotation reveals a finite modulation $A_\gamma \approx 3 \times 10^{-5}$.

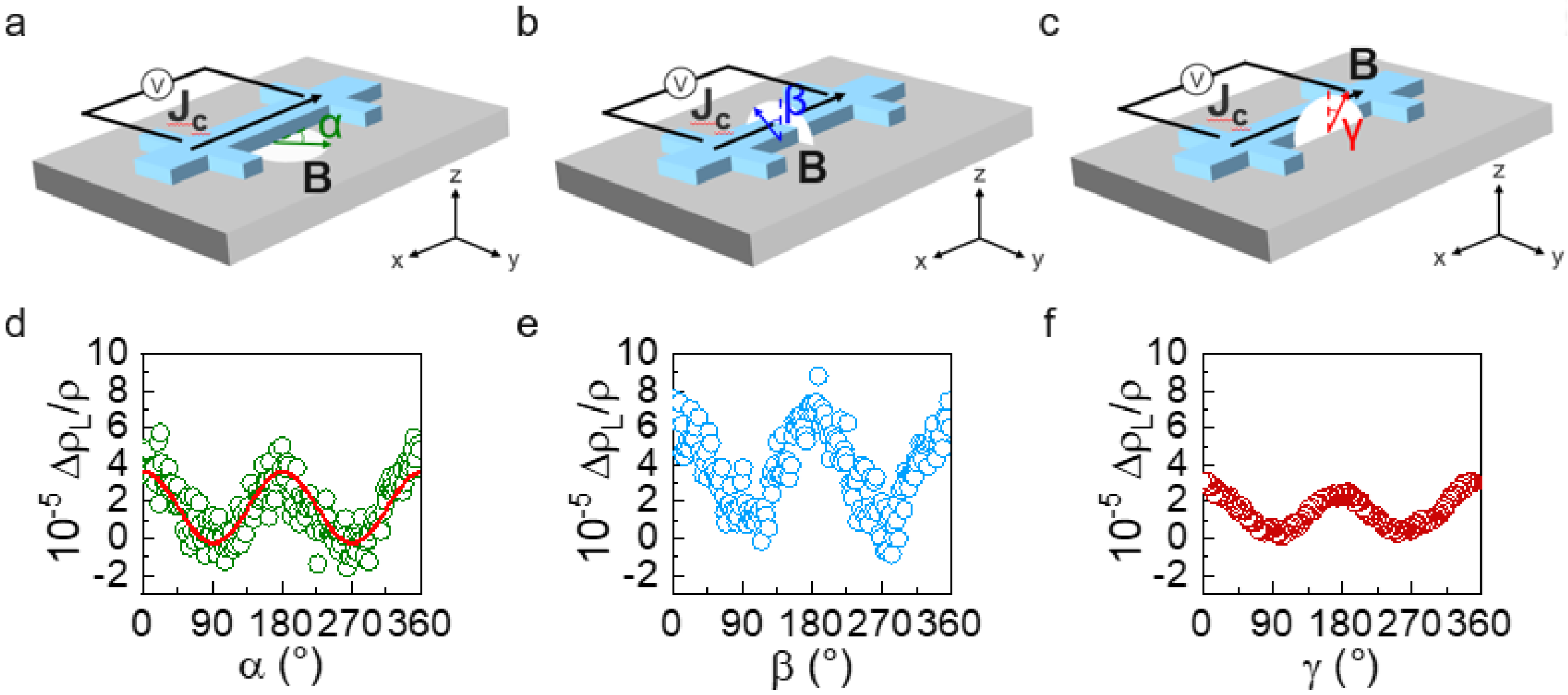

**Fig. 2. Angle-dependent magnetoresistance of LSAT/SVO(20.8 nm) at T = 100 K and B = 9 T.** (a–c) Schematics of the three magnetic-field rotation planes, α (xy), β (yz), and γ (xz), defined with respect to the current direction x. (d–f) Corresponding longitudinal magnetoresistance $\Delta\rho_L/\rho$ as a function of rotation angle. The α-plane data are fitted with the expected $\cos^2$ dependence (red solid line in panel d), yielding $A_\alpha$ = (4.0 ± 0.2) × $10^{-5}$. The α and β rotations display the angular dependence expected for Hanle magnetoresistance, whereas the finite modulation observed in the γ plane indicates an additional non-Hanle contribution.

A γ-plane signal different from zero can in principle arise from a misalignment of the rotation plane. Within the HMR framework, the angular dependence of the longitudinal resistivity is governed by the squared projection of the magnetic-field direction onto the orbital-polarization axis *y*, i.e. $(B_y/B)^2$.[12,24] In an ideal γ-plane rotation, **B** remains in the *xz* plane and therefore $B_y = 0$ for all angles, so no HMR modulation is expected. If the rotation plane is tilted by an angle $\delta$, an unintended field component along *y* appears, with $B_y/B \leq \sin(\delta)$. The resulting spurious modulation is therefore at most $\sin^2(\delta)$ times the Hanle amplitude. Taking the largest measured amplitude, $A_\beta \approx 7 \times 10^{-5}$, as an upper bound, $\delta$ = 5-10° would yield only $5 \times 10^{-7}$ - $2 \times 10^{-6}$, one to two orders of magnitude below the observed γ-plane signal. Reproducing the measured amplitude of $\approx 3 \times 10^{-5}$ would require $\delta \approx 40°$, far beyond our alignment accuracy. Geometrical misalignment therefore cannot account for the γ-plane modulation, which indicates an additional magnetoresistance contribution.[12]

Let $\rho_i$ denote the longitudinal resistivity with the magnetic field applied along direction i. With all three rotations referenced to their 90° value, the measured amplitudes are differences between the same three quantities, $A_\alpha = \rho_x - \rho_y$, $A_\beta = \rho_z - \rho_y$, and $A_\gamma = \rho_z - \rho_x$, and therefore satisfy $A_\gamma = A_\beta - A_\alpha$ identically. Our data obey this relation within the experimental noise, with $A_\beta - A_\alpha \approx 3.0 \times 10^{-5}$, in agreement with the measured $A_\gamma \approx 3 \times 10^{-5}$, confirming the mutual consistency of the three rotation geometries. Because this relation is geometric, it does not by itself identify the microscopic origin of the magnetoresistance. Within an HMR-based decomposition, the inequality between $A_\alpha$ and $A_\beta$ can be described by an additional contribution superimposed on an HMR-like response. For L ∥ y, the ideal HMR contribution is the same for B ∥ x and B ∥ z and vanishes for B ∥ y. We further consider an additional contribution $A_{add}$ that vanishes for B ∥ J and is the same for B ⊥ J, as expected for ordinary Lorentz magnetoresistance. Denoting by H the HMR-like amplitude, this assumption gives $A_\gamma = A_{add}$, $A_\beta = H$ and $A_\alpha = H - A_{add}$. The measured values, $H \approx A_\beta \approx 7 \times 10^{-5}$ and $A_{add} \approx A_\gamma \approx 3 \times 10^{-5}$, are consistent with $A_\alpha \approx 4 \times 10^{-5}$ and imply that the α-plane amplitude underestimates the HMR-like contribution. A similar but much smaller (≈10%) extra contribution was reported for vanadium,[12] where the corresponding B ∥ z component was negative and was attributed to weak localization. In SVO, by contrast, the additional **B** ∥ z contribution is positive, which does not support a WL origin (see Discussion).

### *Field dependence of the magnetoresistance*

To further discriminate the HMR response from additional magnetoresistance contributions, we measured the longitudinal magnetoresistance as a function of magnetic field applied along the three principal directions (x, y, and z) of the Hall bar (Fig. 3a). For **B** ∥ x and **B** ∥ z, the magnetic field is transverse to the orbital angular-momentum polarization **L** ∥ y generated by the OHE, whereas for **B** ∥ y it is parallel to **L**. The longitudinal magnetoresistance is negative for all three field orientations. At 9 T, $\Delta\rho_L/\rho$ reaches approximately $-4 \times 10^{-5}$ for **B** ∥ x, $-7 \times 10^{-5}$ for **B** ∥ y, and $-2 \times 10^{-5}$ for **B** ∥ z. The **B** ∥ z curve additionally shows a weak upturn above $B_z \approx 6$ T, indicative of an additional positive high-field contribution in the out-of-plane configuration. The symmetry expected for HMR is more clearly revealed by considering differences between the longitudinal resistivities measured for the three field orientations. Using the notation introduced above, we define $\Delta\rho_{x-y}(B) \equiv \rho_x(B) - \rho_y(B)$ and $\Delta\rho_{z-y}(B) \equiv \rho_z(B) - \rho_y(B)$. Since the HMR contribution vanishes for **B** ∥ y, these differences suppress the magnetoresistance component common to the different field orientations while retaining the HMR contribution.

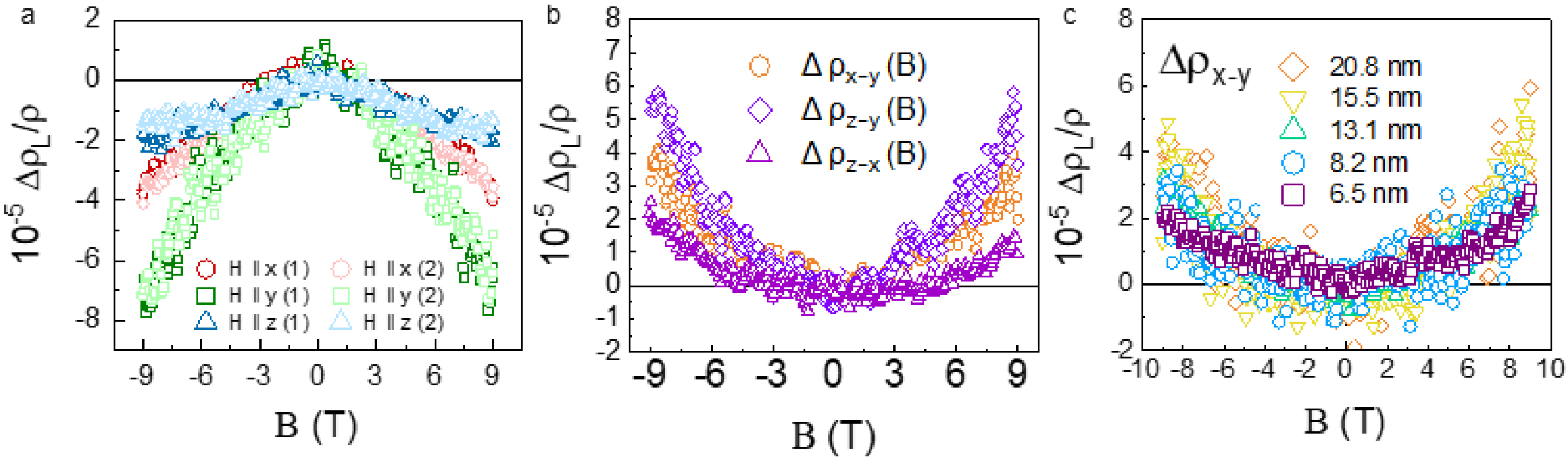

**Fig. 3. Field-dependent magnetoresistance of SVO films at T = 100 K.** (a) Longitudinal magnetoresistance $\Delta\rho_L/\rho$ of the LSAT/SVO(20.8 nm) film for magnetic fields applied along the x, y, and z directions. Two independent field sweeps (B1 and B2) are shown for each orientation, confirming the reproducibility of the measurements. (b) Field-dependent resistivity differences $\Delta\rho_{x-y}/\rho$ and $\Delta\rho_{z-y}/\rho$, where $\Delta\rho_{x-y} \equiv \rho_x - \rho_y$ and $\Delta\rho_{z-y} \equiv \rho_z - \rho_y$. Both differences exhibit a positive, even-in-field response with a similar field evolution, consistent with the expected HMR contribution. Their small deviation at high fields indicates an additional non-Hanle magnetoresistance contribution that depends on the field orientation, most evident for B ∥ z. (c) $\Delta\rho_{x-y}/\rho$ for the broader SVO film series, showing a reproducible even-in-field HMR-like response across samples with different thicknesses and longitudinal resistivities. Because the resistivity varies substantially across the series, these data are used as a robustness test of the HMR phenomenology rather than as a resistivity-uniform thickness series for extracting the orbital diffusion length.

Indeed, as shown in Fig. 3b, both resistivity differences display a positive, even-in-field response with similar, predominantly parabolic field dependences that overlap at low field and separate at high field. The positive, approximately quadratic $\Delta\rho_{x-y}/\rho$ response is the principal field-dependent signature consistent with HMR. Fig. 3b also shows $\Delta\rho_{z-x} = \Delta\rho_{z-y} - \Delta\rho_{x-y}$, in which an ideal HMR contribution cancels; this curve is not independent of the other two, but directly displays the orientation-dependent background. The residual deviation between $\Delta\rho_{x-y}/\rho$ and $\Delta\rho_{z-y}/\rho$ at high fields indicates an additional orientation-dependent magnetoresistance contribution, most evident for B ∥ z. The positive contribution observed for B ∥ z is compatible with an ordinary Lorentz-like magnetoresistance. The field dependences are therefore consistent with an HMR contribution coexisting with additional magnetoresistance backgrounds. Because the microscopic origin and possible anisotropy of the negative background are not independently established, a residual contribution to $\Delta\rho_{x-y}$ cannot be completely excluded.

The field-dependent and angular-dependent measurements probe the same principal-axis resistivity differences in complementary ways. At 9 T the field sweeps give $\Delta\rho_{x-y}/\rho \approx 4 \times 10^{-5}$, comparable with $A_\alpha = (4.0 \pm 0.2) \times 10^{-5}$ from the α-plane rotation, while $\Delta\rho_{z-y}/\rho \approx 6 \times 10^{-5}$ is comparable with $A_\beta \approx 7 \times 10^{-5}$. This quantitative consistency supports the reproducibility of the measured anisotropic magnetoresistance and argues against drift or instrumental offsets, while the field sweeps provide the additional information on its magnetic-field dependence.

To test whether the HMR phenomenology observed in the 20.8 nm film persists across the broader film series, $\Delta\rho_{x-y}/\rho$ was also measured for all investigated thicknesses (Fig. 3c). All samples display a positive, even-in-field response with a similar characteristic field evolution despite substantial differences in thickness and longitudinal resistivity, indicating that the HMR-like field response is reproducible across the SVO series. The complete angular- and field-dependent datasets are reported in Supplementary Figs. S5 and S6, together with the sample-specific transport and structural parameters in Supplementary Table S1.

### ***HMR framework and orbital transport parameters***

To quantify the orbital response of the 20.8 nm film, we use the standard diffusive HMR framework.[12,24] At T = 100 K and B = 9 T, the α-plane angle-dependent magnetoresistance gives $A_\alpha = (3.96 \pm 0.19) \times 10^{-5}$ (Fig. 2d), with $\rho(100\ \mathrm{K}) = 15.6\ \mu\Omega\cdot\mathrm{cm}$. Within this framework, the longitudinal normalized resistivity change at magnetic field B is described by Eqs. (1) - (2).

$$A_\alpha \equiv \Delta\rho_L(B)/\rho = 2\,\theta_{OH}^2\,\{(\lambda_{OD}/t)\tanh[t/(2\lambda_{OD})] - \mathrm{Re}[(\Lambda/t)\tanh(t/2\Lambda)]\} \quad (1)$$

$$\Lambda^{-2} = \lambda_{OD}^{-2} + i\,g\,\mu_B\,B/(D_O\,\hbar) \quad (2)$$

where $t$ is the film thickness, $\lambda_{OD}$ is the orbital diffusion length, $D_O$ is the orbital diffusion coefficient, g is the Landé factor, $\mu_B$ is the Bohr magneton, and Λ is the complex field-dependent diffusion length. The orbital Hall angle and orbital Hall conductivity are related by $\theta_{OH} = \sigma_{OH}\rho$, with ρ evaluated at the measurement temperature. Following the convention of Ref.[12], whose equations we adopt, this relation defines $\sigma_{OH}$ in units of $(\hbar/2e)\ \Omega^{-1}\mathrm{cm}^{-1}$; values quoted below in units of $(\hbar/e)\ \Omega^{-1}\mathrm{cm}^{-1}$, for direct comparison with the first-principles results, are therefore smaller by a factor of two. The orbital relaxation time is $\tau_{OD}$ $= \lambda_{OD}^2/D_O$. The diffusive HMR model underlying Eqs. (1) - (2) is described in Refs.[12,24], while details of its application to the present SVO data and the sensitivity analysis with respect to $\lambda_{OD}$ are provided in Supplementary Section 8.

The field dependence of the HMR provides an estimate of the orbital relaxation time. For the 20.8 nm film, fits of the α-plane angular magnetoresistance give $A_\alpha(3\ T) = (4.68 \pm 0.68) \times 10^{-6}$ and $A_\alpha(9\ T) = (3.96 \pm 0.19) \times 10^{-5}$, corresponding to $A_\alpha(3\ T)/A_\alpha(9\ T) = 0.1182 \pm 0.0181$ (Supplementary Fig. S4). We use this ratio because the overall $\theta_{OH}^2$ prefactor is common to both fields and therefore cancels, leaving the field dependence controlled by $\tau_{OD}$. Evaluating Eqs. (1) - (2) for t = 20.8 nm, $\lambda_{OD}$ = 2 nm and g = 2 gives a best estimate $\tau_{OD}$ = 0.20 ps. Because the measured ratio is compatible with the quadratic low-field value $(3/9)^2 = 0.111$, arbitrarily short relaxation times remain consistent with the data and no finite lower bound on $\tau_{OD}$ can be set, whereas the upper edge of the experimental ratio corresponds to $\tau_{OD} \approx 0.39$ ps. The full derivation is given in Supplementary Section 8.

The near-quadratic field dependence is independently confirmed by the field sweeps. Fitting the $\Delta\rho_{x-y}/\rho$ response of the 20.8 nm film shown in Fig. 3b as $CB^2$ gives $C = (5.93 \pm 0.59) \times 10^{-7}\ T^{-2}$. Since C scales approximately as $\theta_{OH}^2\tau_{OD}^2$ in the low-field regime, it cannot by itself determine $\tau_{OD}$.

For comparison, orbital-HMR measurements in Mn report $\tau_{OD}$ values of about 1–2.5 ps depending on crystalline order, several times longer than the present best estimate of 0.20 ps for SVO.[11] Directly comparable HMR-derived orbital relaxation times have not yet been reported for conducting oxides, including $SrRuO_3$. The comparatively short $\tau_{OD}$ found in SVO may reflect its narrow $t_{2g}$ bandwidth and distinct orbital hybridization.

To estimate the orbital Hall response without relying on the precise value of $\tau_{OD}$, we use the saturation limit of Eq. (1). As $B \rightarrow \infty$, the field-dependent term vanishes and the HMR amplitude approaches its maximum value. The measured finite-field amplitude must therefore satisfy $A_\alpha \leq 2\theta_{OH}^2(\lambda_{OD}/t)\tanh[t/(2\lambda_{OD})]$. Using $\lambda_{OD}$ = 2 nm, t = 20.8 nm and $A_\alpha(9\ T) = (3.96 \pm 0.19) \times 10^{-5}$ gives the conservative lower-bound estimate $\theta_{OH,LB} = 0.0144 \pm 0.0004$, and hence $\sigma_{OH,LB} = (460 \pm 11)\ (\hbar/e)\ \Omega^{-1}cm^{-1}$. Varying the reference $\lambda_{OD}$ between 0.5 and 5 nm gives $\sigma_{OH,LB} = (920 \pm 22)$ to $(296 \pm 7)\ (\hbar/e)\ \Omega^{-1}cm^{-1}$ (Supplementary Section 8). Since the non-Hanle contribution reduces the α-plane amplitude (see above), the estimate is conservative also in this respect. For $\lambda_{OD}$ = 2 nm, this lower bound is already comparable to the intrinsic DFT value $\sigma_{OH}^{DFT} = 390$ $(\hbar/e)\ \Omega^{-1}cm^{-1}$.

### ***Thickness and conductivity dependence of the orbital response***

We next compare the complete SVO series. Figure 4a summarizes the amplitude $A_\alpha$ extracted from the $\cos^2(\alpha)$ signal at 100 K and 9 T. $A_\alpha$ increases monotonically from $(1.59 \pm 0.04) \times 10^{-5}$ for the 6.5 nm film to $(4.0 \pm 0.2) \times 10^{-5}$ for the 20.8 nm film. In contrast, the longitudinal resistivity is non-monotonic across the same series (Supplementary Table S1). Because thickness, conductivity and potentially $\lambda_{OD}$ vary simultaneously, the thickness dependence is not fitted to extract an orbital diffusion length. To compare all films on the same basis, we apply the HMR framework to all films using the same reference $\lambda_{OD}$ = 2 nm. This gives $\sigma_{OH,LB}$ = (52.8 ± 0.7), (179 ± 6), (153 ± 3), (194 ± 4) and (460 ± 11) (ħ/e) $\Omega^{-1}$cm$^{-1}$ for thicknesses 6.5, 8.2, 13.1, 15.5 and 20.8 nm, respectively. The quoted uncertainties propagate the formal $A_\alpha$ fit errors. Overall, $\sigma_{OH,LB}$ increases with $\sigma_{xx}$ (Fig. 4b). Because $\lambda_{OD}$ may itself vary with conductivity, we treat this as an empirical correlation rather than a scaling law for the intrinsic $\sigma_{OH}$, because separating conductivity and orbital relaxation effects would require a dedicated disorder-controlled series.

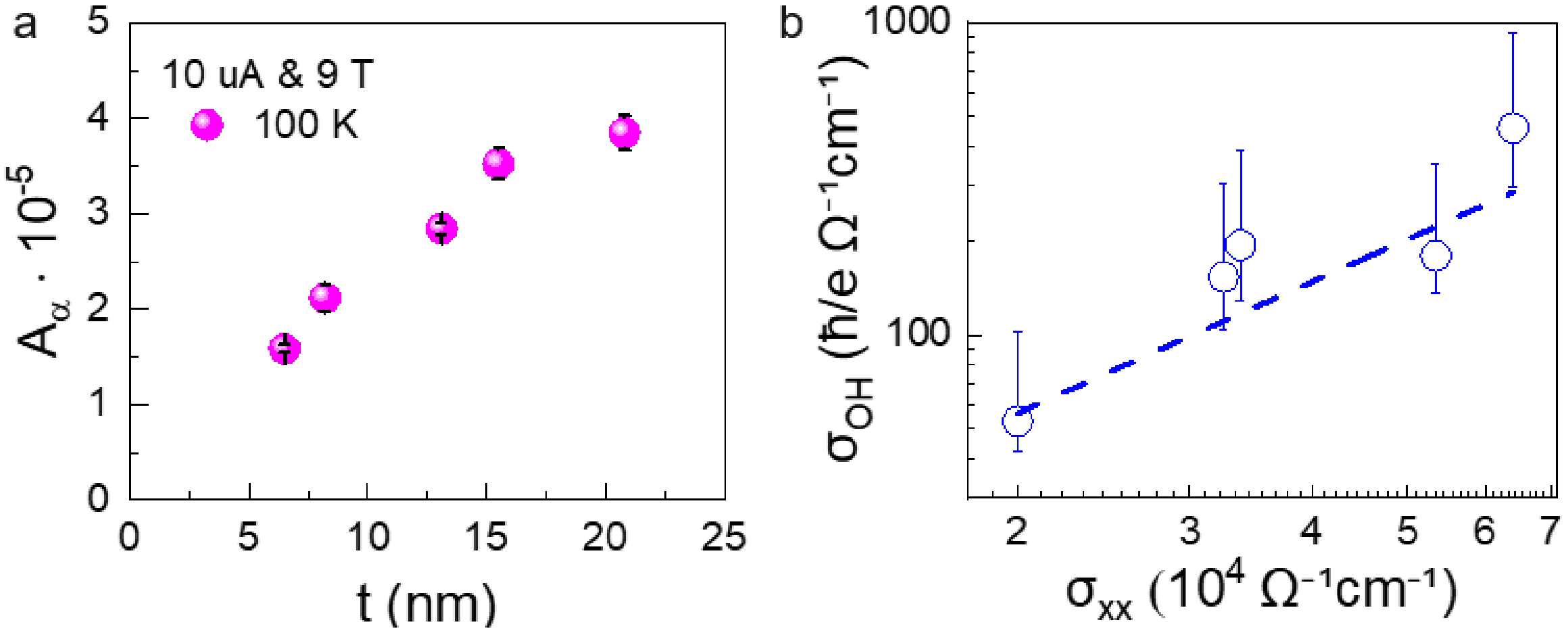

**Figure 4. Thickness and conductivity dependence of the orbital response in SVO films.** (a) α-plane magnetoresistance amplitude at T = 100 K and B = 9 T as a function of film thickness; error bars are the formal $\cos^2$-fit uncertainties. (b) Saturation-limit lower-bound estimates $\sigma_{OH,LB}$, calculated using the same reference $\lambda_{OD}$ = 2 nm for all films, plotted as a function of the longitudinal conductivity $\sigma_{xx} = 1/\rho(100\text{ K})$. Vertical bars in panel (b) represent the model sensitivity to $\lambda_{OD}$ varied between 0.5 and 5 nm. The dashed line is a guide to the eye.

### *First-principles orbital Hall response*

To establish an independent material-specific benchmark for the experimentally extracted orbital response, first-principles calculations of the intrinsic orbital and spin Hall responses were performed for bulk SVO at zero strain, used here as the intrinsic reference state. Figure 5a shows the calculated electronic band structure, with the states around the Fermi level predominantly derived from the V $t_{2g}$ manifold.[14,15] The corresponding energy-dependent orbital Hall conductivity and spin Hall conductivity are shown in Fig. 5b. At the Fermi level, the calculations yield $\sigma_{OH}^{DFT}$ = +390 (ħ/e)(Ω cm)$^{-1}$, whereas $\sigma_{SH}^{DFT}$ = −12 (ħ/e)(Ω cm)$^{-1}$. The DFT value should be regarded as an intrinsic clean-limit benchmark rather than as the conductivity necessarily expected experimentally. Tang and Bauer showed that disorder-induced vertex corrections can strongly modify $\sigma_{OH}$, particularly when the states at the Fermi level involve substantial s-d hybridization.[16] Moreover, the quantum-kinetic treatment of Valet et al.[17] shows that the orbital response relevant to transport and edge accumulation can differ substantially from the conventional intrinsic Kubo orbital Hall conductivity. Its relation to the measured lower bound is discussed below. The calculated transverse angular-momentum response of SVO is strongly orbital dominated, with $|\sigma_{OH}^{DFT}/\sigma_{SH}^{DFT}| \approx 33$ at the Fermi level. Computational details are provided in Supplementary Section 7. Table 1 summarizes the orbital transport parameters estimated for the 20.8 nm SVO film and compares them with the experimental and theoretical values reported for V.[12]

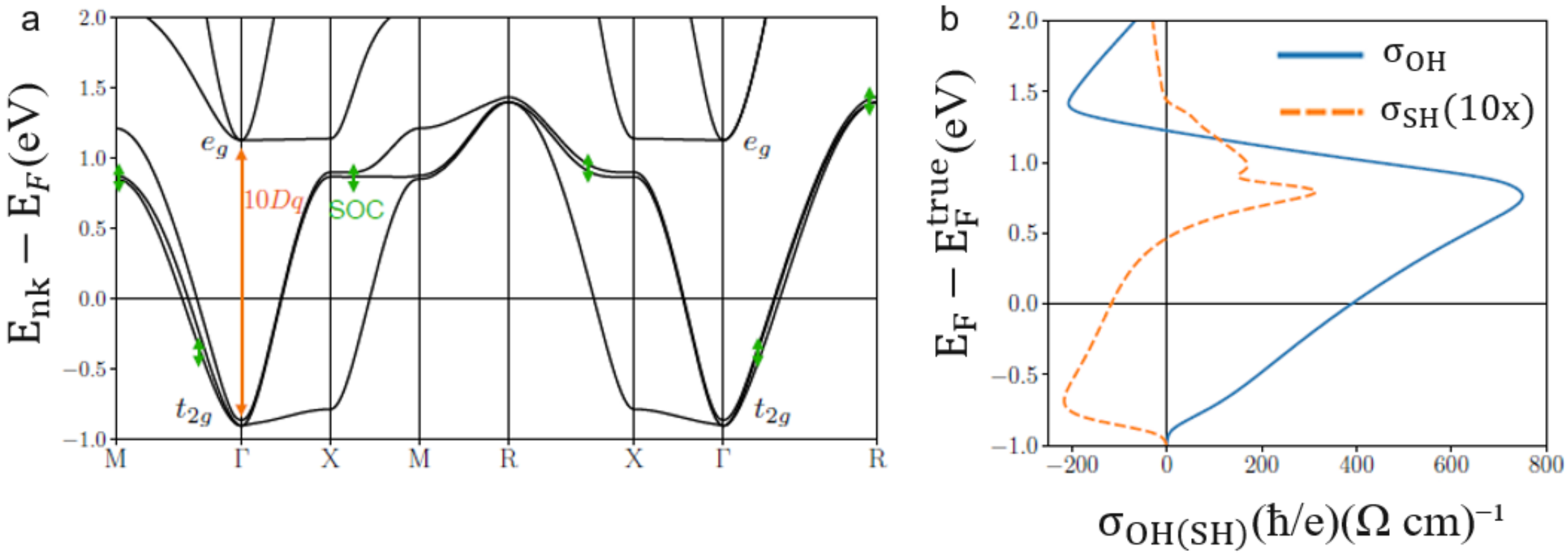

**Figure 5. First-principles electronic structure and intrinsic Hall conductivities of SVO.** (a) Calculated electronic band structure of bulk SVO at zero strain. The crystal-field-split V $t_{2g}$ and $e_g$ manifolds are indicated, together with the characteristic crystal-field energy scale 10Dq and SOC. (b) Energy dependence of the intrinsic orbital Hall conductivity (solid blue line) and spin Hall conductivity (dashed orange line; multiplied by 10 for visibility), with energy referenced to the Fermi level $E_F$. At $E_F$, the calculated conductivities are $\sigma_{OH}^{DFT}$ = +390 (ħ/e)(Ω cm)$^{-1}$ and $\sigma_{SH}^{DFT}$ = −12 (ħ/e)(Ω cm)$^{-1}$, demonstrating a predominantly orbital transverse response.

**Table 1.** Comparison of orbital transport parameters estimated from HMR for SVO (this work, representative LSAT/SVO(20.8 nm) film, 100 K) with experimental and first-principles values for V. For SVO, $\lambda_{OD}$ = 2 nm is used as a reference value. The 3-9 T α-plane field evolution provides the effective $\tau_{OD}$ reported below, while $\theta_{OH}$ and $\sigma_{OH}$ are given as conservative saturation-limit lower-bound estimates. The quoted SVO uncertainties propagate the formal $A_\alpha$ fit errors. Experimental values for V are those of Ref. 12 at 100 K. Orbital Hall conductivities are quoted throughout in units of (ħ/e) Ω$^{-1}$cm$^{-1}$; the value reported for V in the original work, 76 (ħ/2e) Ω$^{-1}$cm$^{-1}$, has been converted accordingly.

| Parameter | SVO (this work) | DFT (SVO) | V (Ref. 12) | DFT (V) |
|---|---|---|---|---|
| ρ(100 K) (μΩ·cm) | 15.6 | — | 264.4 | — |
| $A_\alpha$ ($10^{-5}$) | 3.96 ± 0.19 | — | ≈13 | — |
| $\theta_{OH}$ lower bound | 0.0144 ± 0.0004 | — | 0.020 ± 0.001 | — |
| $\lambda_{OD}$ (nm) | 2 (assumed) | — | 1.8 ± 0.3 | 0.5 |

| Parameter | SVO (this work) | DFT (SVO) | V (Ref. [12]) | DFT (V) |
|---|---|---|---|---|
| $\tau_{OD}$ (ps) | 0.20 (best estimate; upper compatible range ≈ 0.39) | — | — | — |
| $\sigma_{OH}$ lower bound (ħ/e $\Omega^{-1}cm^{-1}$) | 460 ± 11 | ≈ 390 | ≈ 38 | 4500 - 6050 |
| $\sigma_{OH}^{exp}$ / $\sigma_{OH}^{DFT}$ (lower bound, $\lambda_{OD}$ = 2 nm) | 1.18 ± 0.03 | 1 | ≈ 0.006–0.008 | 1 |

## Discussion

**Comparison with pure 3d metals.** For the most conductive film, the lower-bound $\sigma_{OH}$ of SVO reaches the scale of the intrinsic DFT value, whereas the experimental orbital Hall conductivity reported for V is strongly suppressed relative to its intrinsic prediction. Recent theory shows that disorder-induced vertex corrections can strongly suppress $\sigma_{OH}$ in systems with substantial s-d hybridization. The low-energy electronic structure of SVO is instead dominated by V $t_{2g}$-derived states hybridized with O 2p states. Whether this distinct orbital character makes the orbital current less sensitive to vertex corrections remains an open theoretical question and is not assumed here.

Second, the quantum-kinetic formulation of orbital transport in Ref.[17] shows that the nonequilibrium orbital response relevant to transport and edge accumulation can differ from the conventional intrinsic Kubo orbital Hall conductivity. In $d^1$ oxides, the localized and hybridized character of the $t_{2g}$ states and Coulomb correlations may modify the relevant orbital matrix elements. The first-principles calculations presented here therefore provide an intrinsic benchmark, while a microscopic description of the interplay between correlations, scattering and orbital relaxation would require extensions beyond the present calculation.[11,13]

[13]The conductivity dependence of the orbital response provides a complementary perspective on scattering. As noted in the Introduction, Mn and $SrRuO_3$ show contrasting conductivity scaling, indicating that scattering and orbital relaxation affect the OHE in a material-dependent manner. The empirical positive correlation observed in SVO adds a further case, whose separation into orbital Hall and orbital-relaxation contributions requires the disorder-controlled series discussed above.

**Origin of the non-Hanle contribution.** In V, the analogous non-Hanle component was attributed to weak WL, consistent with the resistivity upturn observed in the sputtered films of Ref. [12]. In the 20.8 nm SVO film no resistivity upturn is observed down to 2 K (Fig. 1c). This alone does not exclude WL: for a sheet conductance $t/\rho_0 \approx 0.15$ S, a quantum correction of order $e^2/(2\pi^2\hbar) \approx 1.2 \times 10^{-5}$ S corresponds to a relative change of only ~$10^{-4}$, which would be difficult to resolve in $\rho(T)$. An isotropic WL-like negative magnetoresistance may therefore contribute to the common background in Fig. 3a, but it cancels in the resistivity differences. WL is instead unlikely to account for the anisotropic non-Hanle contribution: a two-dimensional WL correction is largest for out-of-plane fields and would make $A_\gamma = \rho_z - \rho_x$ negative, as reported for V, whereas the measured $A_\gamma$ is positive.

Possible alternative origins include: (i) electron-electron interaction corrections in a quasi-2D diffusive regime, which can produce field-dependent magnetoresistance;[25] (ii) correlation-driven scattering involving spin and orbital fluctuations, which has been predicted to modify quasiparticle lifetimes in ultrathin SVO;[26] (iii) band-geometric magnetoresistance associated with field-induced Berry-curvature effects in centrosymmetric metals;[27] and (iv) ordinary Lorentz magnetoresistance, which is largest for $B \perp J$ and therefore naturally produces a modulation in the γ plane. In the α plane, the latter has the opposite sign to HMR: Lorentz magnetoresistance raises ρ for $B \perp J$ ($\alpha = 90°$), whereas HMR raises ρ for $B \perp L$ ($\alpha = 0°$). The measured α-plane amplitude is positive, $A_\alpha = +4.0 \times 10^{-5}$. Therefore, a Lorentz-like contribution with the sign inferred from the γ-plane response subtracts from the HMR amplitude rather than generating it, making the $\theta_{OH}$ extracted from the α-plane modulation a conservative lower-bound estimate.[10,11]

The relatively large γ-plane modulation, amounting to 76% of the α-plane and 43% of the β-plane amplitude, also differs from the nearly isotropic orbital-HMR response reported in Mn, where no comparably large non-Hanle component was observed.10,11 This suggests that the additional contribution is related to the electronic structure of SVO rather than to the HMR geometry. Its microscopic origin remains unresolved and constitutes the main uncertainty in quantitatively isolating the HMR contribution.

**Implications for orbital electronics.** If our results are confirmed by independent techniques and replicated in other $d^1$ transition metal oxides, they would suggest that the family of perovskite metals whose Fermi level lies within a single $t_{2g}$ manifold (e.g., $SrVO_3$, $CaVO_3$ and $SrNbO_3$) could serve as a versatile materials platform for orbital electronics. In addition to potentially larger $\sigma_{OH}$, these oxides could offer strain-tunable orbital occupancy via substrate selection, which may enable control of the OHE through the orbital populations of the $t_{2g}$ manifold, a degree of freedom that is less accessible in elemental metals.

In a broader experimental context, the lower-bound orbital Hall conductivity obtained for SVO is substantially larger than the values extracted from single-layer HMR in Mn and V, which are in the tens of $(\hbar/e)\ \Omega^{-1}\text{cm}^{-1}$.[10,12] Larger orbital Hall conductivities, of order several $10^3\ (\hbar/e)\ \Omega^{-1}\text{cm}^{-1}$, have been inferred from orbital-torque measurements in Cr-based heterostructures.[28] Conductivity tuning in $SrRuO_3$ has also been used to enhance orbital torque and achieve a threefold reduction in switching power.[13] These torque-based estimates, however, involve interfacial orbital transmission and orbital-to-spin conversion and are therefore not directly comparable with single-layer HMR. Within the directly comparable HMR framework, the orbital Hall angle of SVO ($\theta_{OH,LB} = 0.0144$) is similar to that of V (0.020), so that its larger $\sigma_{OH}$ mainly reflects its low resistivity. This combination of a sizable orbital Hall angle and metallic conductivity supports the potential of SVO as an orbital-current source. Its relevance for orbital-torque devices will ultimately depend on the efficiency of orbital transmission and conversion at interfaces.

## Conclusions

We have observed reproducible magnetoresistance signatures in epitaxial LSAT/SVO that are consistent with an orbital-HMR contribution, most notably a positive, even-in-field and approximately quadratic $\Delta\rho_{x-y}/\rho$ response. The angular measurements characterize the corresponding magnetoresistance anisotropy and reveal an additional non-Hanle contribution. Within the diffusive HMR framework, the 20.8 nm film gives a best estimate $\tau_{OD}$ = 0.20 ps, with an upper compatible range of ≈ 0.39 ps for $\lambda_{OD}$ = 2 nm, and saturation-limit lower-bound estimates $\theta_{OH,LB}$ = 0.0144 ± 0.0004 and $\sigma_{OH,LB}$ = (460 ± 11) (ħ/e) $\Omega^{-1}cm^{-1}$. Across the broader SVO series, the HMR-like field response is reproducible and the extracted orbital response varies with transport conditions. Together with the predominantly orbital first-principles response, these results support the presence of orbital HMR in a narrow-band $d^1$ metallic oxide and suggest that $d^1$ perovskite metals are a promising platform for orbital transport. Separating the HMR and non-Hanle contributions quantitatively, for example through temperature- and disorder-dependent studies, is a natural next step.

## Funding Declaration

E.L. and J.F. acknowledge the financial support of the Spanish Ministry of Science and Innovation through Projects PID2023-152225NB-I00 and Severo Ochoa MATRANS42 (CEX2023-001263-S), supported by MICIU/AEI/10.13039/501100011033 and FEDER, EU. F.C. acknowledges funding from MICIU/AEI/10.13039/501100011033 (Grants No. CEX2020-001038-M and CEX2025-001634-M) and from MICIU/AEI and ERDF/EU (Project No. PID2024-155708OB-I00). M.X.A.-P. acknowledges support from MCIU/AEI and ESF Investing in your Future (Fellowship No. PRE-2019–089833).

## Contribution

J.F. and E.L. conceived the experiment. E.L. grew and patterned the SVO thin films and performed their structural and chemical characterization. J.B., M.X.A.-P., and E.L. performed the magnetotransport measurements. J.B. and E.L. analyzed the data. D.G. performed the first-principles calculations of the SVO band structure and the relevant transport parameters. F.C. supervised the activity at CIC nanoGUNE. All authors discussed the results and revised the manuscript. E.L. wrote the manuscript with contributions from all co-authors and led and supervised the project.

## Orbital angular momentum accumulation in $SrVO_3$ thin films

Julien Brehin[1], Montserrat X. Aguilar-Pujol[1], Dongwook Go[2,3], F. Casanova[1], J. Fontcuberta[4], E. Longo[4*]

*1. CIC nanoGUNE BRTA, E-20018 Donostia-San Sebastian, Spain*

*2. Department of Physics, Korea University, Seoul 02841, Republic of Korea*

*3. Center for Quantum Dynamics of Angular Momentum, Pohang 37673, Korea*

*4. Institut de Ciencia de Materials de Barcelona (ICMAB-CSIC), Campus de la UAB, Bellaterra (Barcelona), Spain*

Corresponding author: * *elongo@icmab.es*

### *Supplementary Information*

**Table S1.** Summary of structural, longitudinal-transport and HMR parameters of the $SrVO_3$ (SVO) film series. Film thicknesses t were obtained by XRR except for t = 8.2 nm (marked *), estimated from the growth calibration. ρ and $|A_\alpha|$ are evaluated at 100 K; θOH and σOH are illustrative fixed-$\lambda_{OD}$ estimates using $\lambda_{OD}$ = 2 nm (Section 8). The listed SVO(002) peak positions and derived c and $\varepsilon_{zz}$ values are approximate and are reported only where the film-related diffraction feature is reliably resolved. $\varepsilon_{zz}$ denotes the out-of-plane lattice strain relative to bulk SVO, not the in-plane epitaxial strain relative to LSAT.

| **Nominal t (nm)** | **t (nm)** | **ρ(100 K) (μΩ·cm)** | **$2\theta_{002}$ (°)** | **c (Å)** | **$\varepsilon_{zz}$ (%)** | **$\lvert A_\alpha \rvert$ ($10^{-5}$)** | **$\theta_{OH}$** | **$\sigma_{OH}$ (ħ/e $\Omega^{-1}cm^{-1}$)** |
|---|---|---|---|---|---|---|---|---|
| 5 | 6.5 | 50.0 | — | — | — | 1.59 | 0.00528 | 53 |
| 10 | 8.2* | 18.7 | — | — | — | 2.12 | 0.00670 | 179 |
| 15 | 13.1 | 30.8 | 47.38 | 3.834 | −0.20 | 2.70 | 0.00942 | 153 |
| 20 | 15.5 | 29.6 | 47.35 | 3.837 | −0.14 | 3.40 | 0.01148 | 194 |
| 25 | 20.8 | 15.6 | 47.17 | 3.850 | +0.22 | 3.96 | 0.01435 | 460 |

### 1. Structural characterization of the SVO film series

Figure S1 summarizes the X-ray structural characterization of the SVO series. XRR fits give t = 6.5, 13.1, 15.5, and 20.8 nm for the nominal 5, 15, 20, and 25 nm samples, while t = 8.2 nm for the nominal 10 nm film was estimated from the growth calibration. In the θ–2θ scans, the film-related SVO(002) feature is identified at approximately $2\theta_{002}$ = 47.38°, 47.35°, and 47.17° for the 13.1, 15.5, and 20.8 nm films, respectively. For the 6.5 nm film the SVO(002) contribution is too weak and broadened for a reliable peak-position determination; no θ–2θ scan is available for the 8.2 nm sample. For the (002) reflection, $d_{002} = c/2$ and Bragg's law gives $c = \lambda/\sin\theta$, with $\theta = (2\theta_{002})/2$ and λ = 1.5406 Å. The out-of-plane strain is $\varepsilon_{zz} = (c - c_{bulk})/c_{bulk} \times 100$, using cbulk = 3.842 Å (bulk SVO(002): 2θ ≈ 47.28°). Because only symmetric (002) scans are used, these values quantify only the out-of-plane lattice distortion relative to bulk SVO and do not determine the in-plane lattice parameter, the full tetragonal distortion c/a, the strain relative to LSAT, or the degree of epitaxial relaxation.

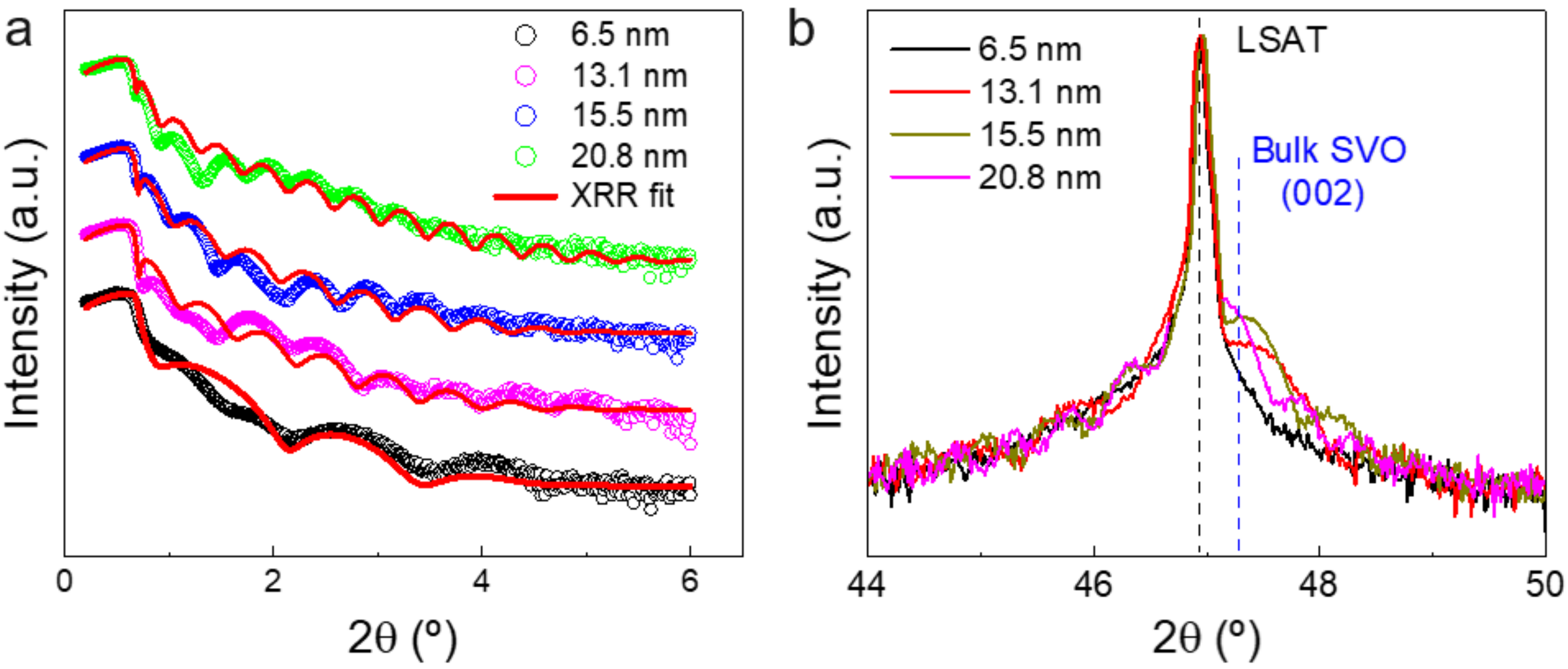

**Figure S1.** X-ray structural characterization of the SVO film series. (a) X-ray reflectivity curves for the four samples for which XRR measurements were available; solid red curves are fits used to determine t. The nominal 10 nm sample (t = 8.2 nm, estimated by interpolation) is not shown. (b) θ–2θ scans around the LSAT(002) and SVO(002) reflections for the 6.5, 13.1, 15.5, and 20.8 nm films. Vertical dashed markers indicate the approximate film-related peak positions together with the bulk SVO(002) reference at 2θ ≈ 47.28°. The 6.5 nm film contribution is not sufficiently resolved for a reliable peak-position determination.

## 2. Longitudinal and Hall transport across the SVO film series

All investigated films remain metallic over the measured temperature range, although the absolute resistivity varies substantially from sample to sample (Fig. S2a). The film-thickness values listed in Table S1 were used for all resistivity conversions. For a Hall bar of length l, width W, and film thickness t, the longitudinal resistivity was calculated from the measured four-probe resistance $R_{xx}$ as

$$\rho_{xx} = R_{xx} \frac{Wt}{l}$$

where l = 100 µm and W = 20 µm for the devices used here. With t in nm and $R_{xx}$ in Ω, the conversion becomes

$$\rho_{xx}\ (\mu\Omega \cdot \mathrm{cm}) = 0.02\ R_{xx}(\Omega)\ t(\mathrm{nm})$$

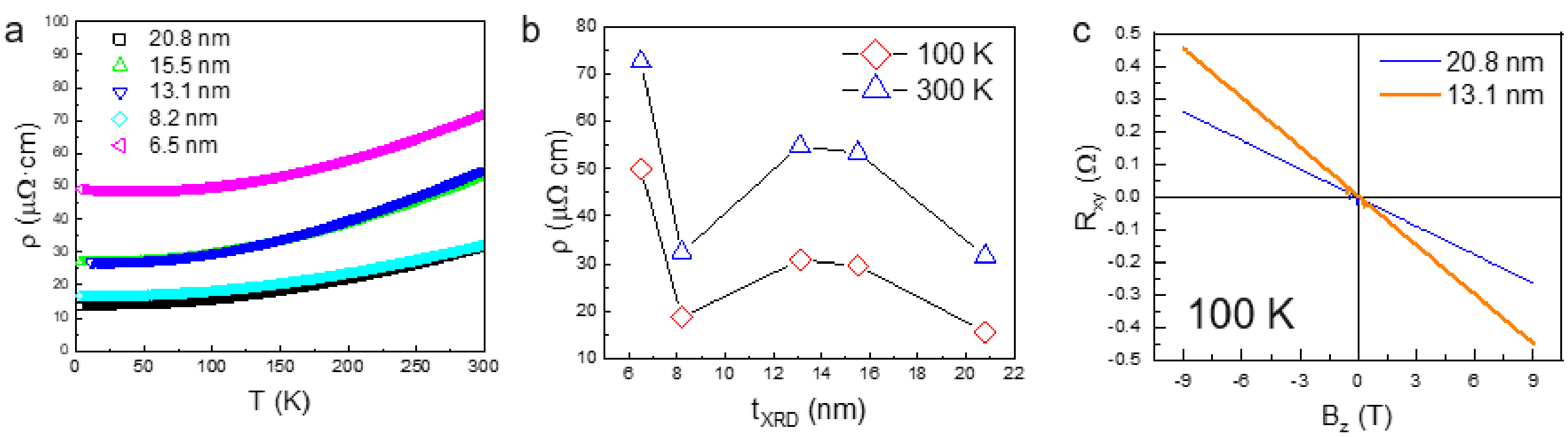

**Figure S2.** Longitudinal and Hall transport overview of the $SrVO_3$ film series. (a) Temperature dependence of the longitudinal resistivity ρ for the five investigated films. (b) ρ at 100 and 300 K as a function of film thickness t; the 8.2 nm point corresponds to the interpolated thickness of the nominal 10 nm film (Table S1). Lines are guides to the eye and are not intended to represent a thickness-dependent fit. (c) Representative transverse Hall resistance Rxy(B) at 100 K for the 13.1 and 20.8 nm patterned films with B ∥ z. The negative Hall slope corresponds to electron-like transport.

Figure S2b shows that the longitudinal resistivity does not evolve monotonically with thickness at either 100 or 300 K. This sample-to-sample variation confirms that the films should not be regarded as a resistivity-uniform thickness series and therefore should not be used as such for a quantitative thickness-dependent extraction of the orbital transport parameters. Figure S2c shows representative Hall measurements performed at 100 K on the 13.1 and 20.8 nm patterned films. In both samples, $R_{xy}$ is linear with the out-of-plane magnetic field and has a negative slope, confirming electron-like transport. The Hall slope $S_H = dR_{xy}/dB$ was obtained from a linear fit, and the three-dimensional Hall coefficient was calculated as

$$R_H = t\,S_H$$

Within a single-band approximation, the effective Hall carrier density and Hall mobility are

$$n_e = \frac{1}{e\,|\,R_H\,|} = \frac{1}{et\,|\,S_H\,|}$$

$$\mu_H = \frac{|\,R_H\,|}{\rho_{xx}} = \frac{1}{n_e e \rho_{xx}}$$

For direct calculation with t in nm, $S_H$ in Ω/T, and $\rho_{xx}$ in μΩ·cm, these expressions become

$$n_e(\mathrm{cm}^{-3}) = \frac{6.2415 \times 10^{21}}{t(\mathrm{nm})\,|\,S_H\,|\,(\Omega/\mathrm{T})}$$

$$\mu_H(\mathrm{cm}^2 \cdot \mathrm{V}^{-1} \cdot \mathrm{s}^{-1}) = \frac{1000\,t(\mathrm{nm})\,|\,S_H\,|\,(\Omega/\mathrm{T})}{\rho_{xx}(\mu\Omega \cdot \mathrm{cm})}$$

A field-independent transverse-voltage offset does not affect the extracted Hall parameters because only the fitted slope enters the calculation. The longitudinal resistivity used to evaluate $\mu_H$ refers to the same patterned device and temperature as the Hall measurement.

**3. X-ray photoemission spectroscopy of the representative LSAT/SVO(20.8 nm) film**

A survey X-ray photoemission spectroscopy (XPS) spectrum was acquired on an air-exposed LSAT/SVO(20.8 nm) film to verify its surface chemical composition and to exclude detectable contamination by common ferromagnetic elements. The spectrum shows the expected Sr, V and O photoemission features, while the binding-energy regions corresponding to the Fe 2p, Co 2p and Ni 2p core levels are indicated and no corresponding peaks are resolved above the survey background. Within the sensitivity of the measurement, the XPS data therefore provide no evidence for Fe-, Co- or Ni-containing contamination.

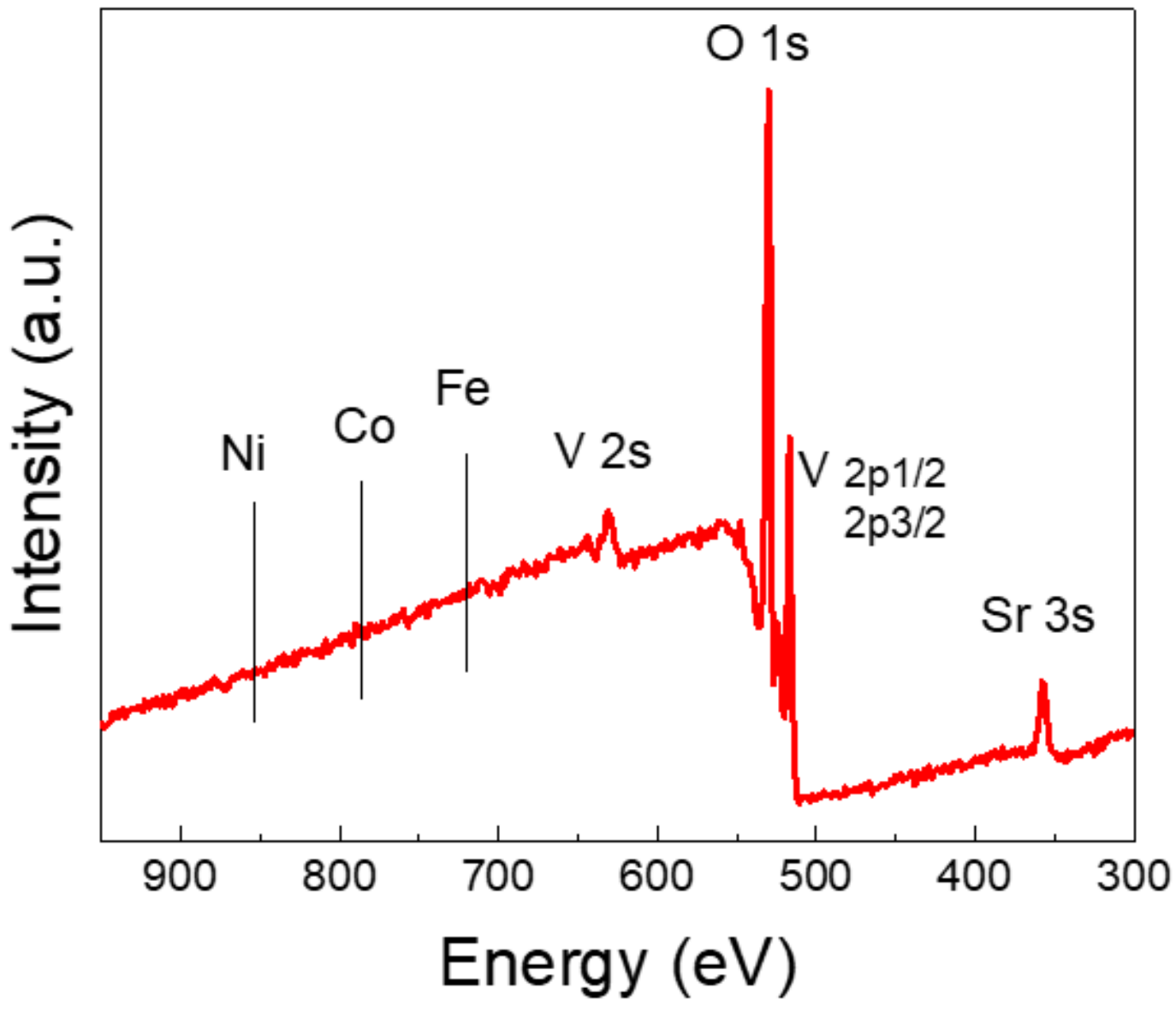

**Figure S3.** Survey X-ray photoemission spectrum of the air-exposed LSAT/SVO(20.8 nm) film. The expected Sr, V and O photoemission features are observed. Vertical markers indicate the binding-energy regions of the Fe 2p, Co 2p and Ni 2p core levels; no corresponding peaks are resolved above the survey background.

### 4. Detailed transport characterization of the representative SVO(20.8 nm) film

Having established the structural and charge-transport behavior of the full film series, we focus here on the 20.8 nm film as a representative sample for the detailed transport and angle-dependent magnetoresistance characterization.

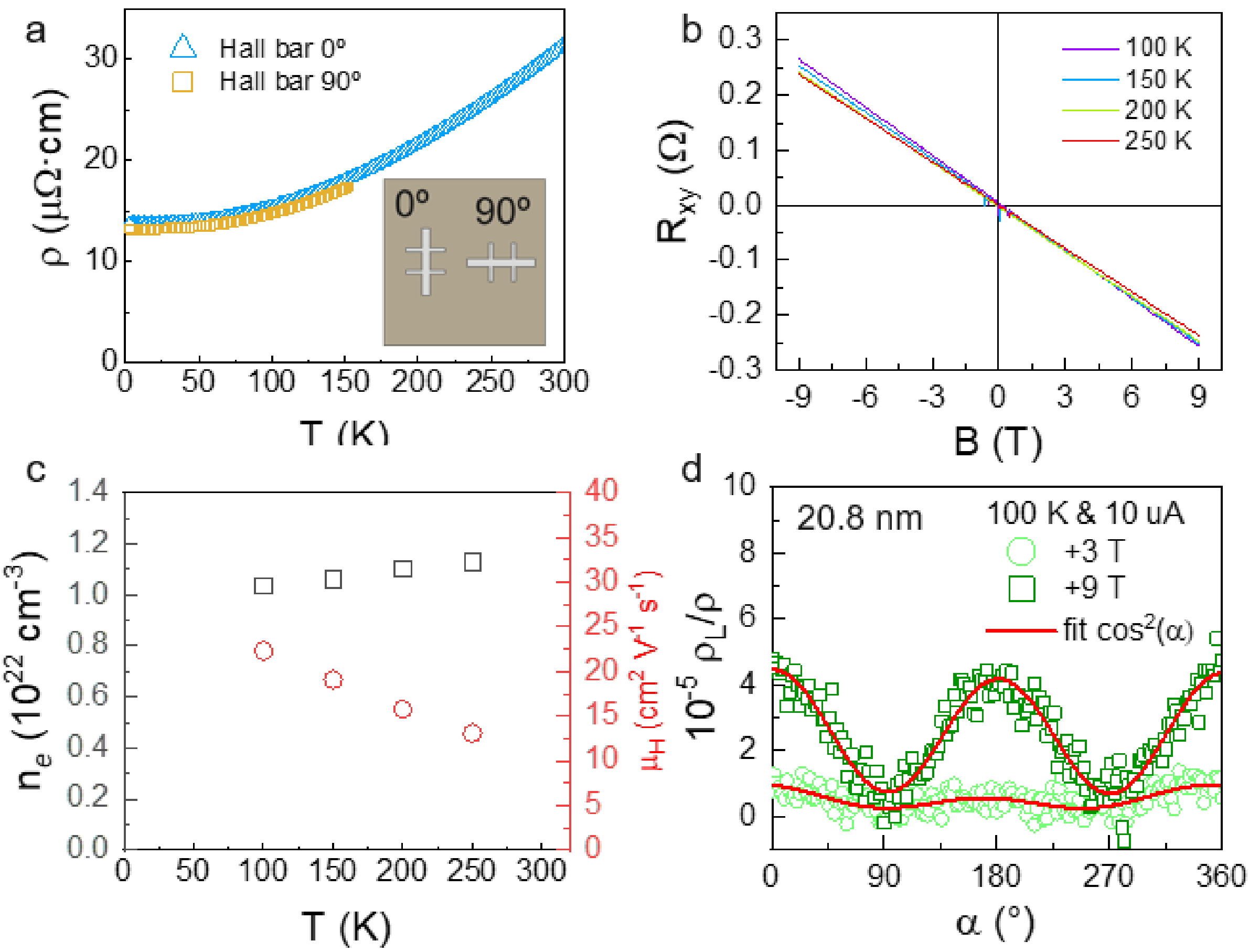

**Figure S4.** Detailed transport characterization of the representative SVO(20.8 nm) film. (a) Temperature dependence of ρ for two mutually perpendicular Hall bars patterned on the same film. (b) Temperature evolution of $R_{xy}(B)$ between 100 and 250 K. (c) Effective single-band Hall carrier density ne and Hall mobility $\mu_H$ extracted over the same temperature range. (d) Field evolution of the α-plane angle-dependent magnetoresistance at 100 K and I = 10 μA for B = 3 and 9 T; solid curves are $\cos^2\alpha$ fits.

Figure S4a compares the longitudinal resistivity of two mutually perpendicular Hall bars patterned on the same 20.8 nm SVO film. The two devices display very similar temperature dependences and comparable absolute resistivities over the complete measured range. Within the experimental resolution, no sizeable static in-plane anisotropy of the longitudinal charge transport is detected. This observation is consistent with the four-fold in-plane structural symmetry and supports that the angle-dependent magnetoresistance is not a trivial consequence of an anisotropic zero-field resistivity.

Figure S4b shows the Hall response of the same representative device between 100 and 250 K. The transverse resistance remains linear in magnetic field throughout this temperature range, with a negative slope at all measured temperatures. Figure S4c summarizes the effective single-band Hall carrier density ne and Hall mobility $\mu_H$. The carrier densities are ne = 1.034, 1.062, 1.104, and 1.130 × $10^{22}$ cm$^{-3}$ at 100, 150, 200, and 250 K, respectively. The corresponding Hall mobilities are $\mu_H$ = 38.7, 32.4, 26.1, and 21.1 cm$^2$ V$^{-1}$ s$^{-1}$. The effective Hall carrier density therefore changes only moderately over this temperature interval, whereas the mobility decreases as the longitudinal resistivity increases. These Hall-derived quantities are used as electronic-transport benchmarks within a single-band description and are not interpreted as a direct measure of the microscopic d-orbital occupation.

Figure S4d compares the α-plane magnetoresistance at B = 3 and 9 T at 100 K. The angular modulation is weak at 3 T and becomes clearly resolved at 9 T, where a $\cos^2\alpha$ response is observed. This field evolution demonstrates that the angular modulation develops with magnetic field rather than representing a field-independent angular background.

**5. Angle-dependent magnetoresistance across the SVO thickness series**

Angle-dependent magnetoresistance was measured in the α rotation plane at T = 100 K and B = 9 T for the complete SVO series. The orbital angular-momentum polarization generated by the OHE is denoted by L and is oriented along the y direction. In this geometry the magnetic field rotates in the film plane between B ∥ x, parallel to the current direction, and B ∥ y, parallel to L (see the measurement geometry in the main text). All investigated films display the same π-periodic $\cos^2$-like modulation, with maxima for B ∥ x and minima for B ∥ y.

The amplitude $|A_\alpha|$ was extracted from a $\cos^2$ fit for each film. Despite the variation of the longitudinal resistivity across the series, the angular response retains the same phase and symmetry for all thicknesses. The fitted amplitudes are summarized as a function of film thickness in Figure S5f and are also reported in Table S1. These data establish the reproducibility of the HMR-like angular response, while the non-uniform resistivity of the series prevents a direct interpretation of the amplitude variation as a pure thickness dependence.

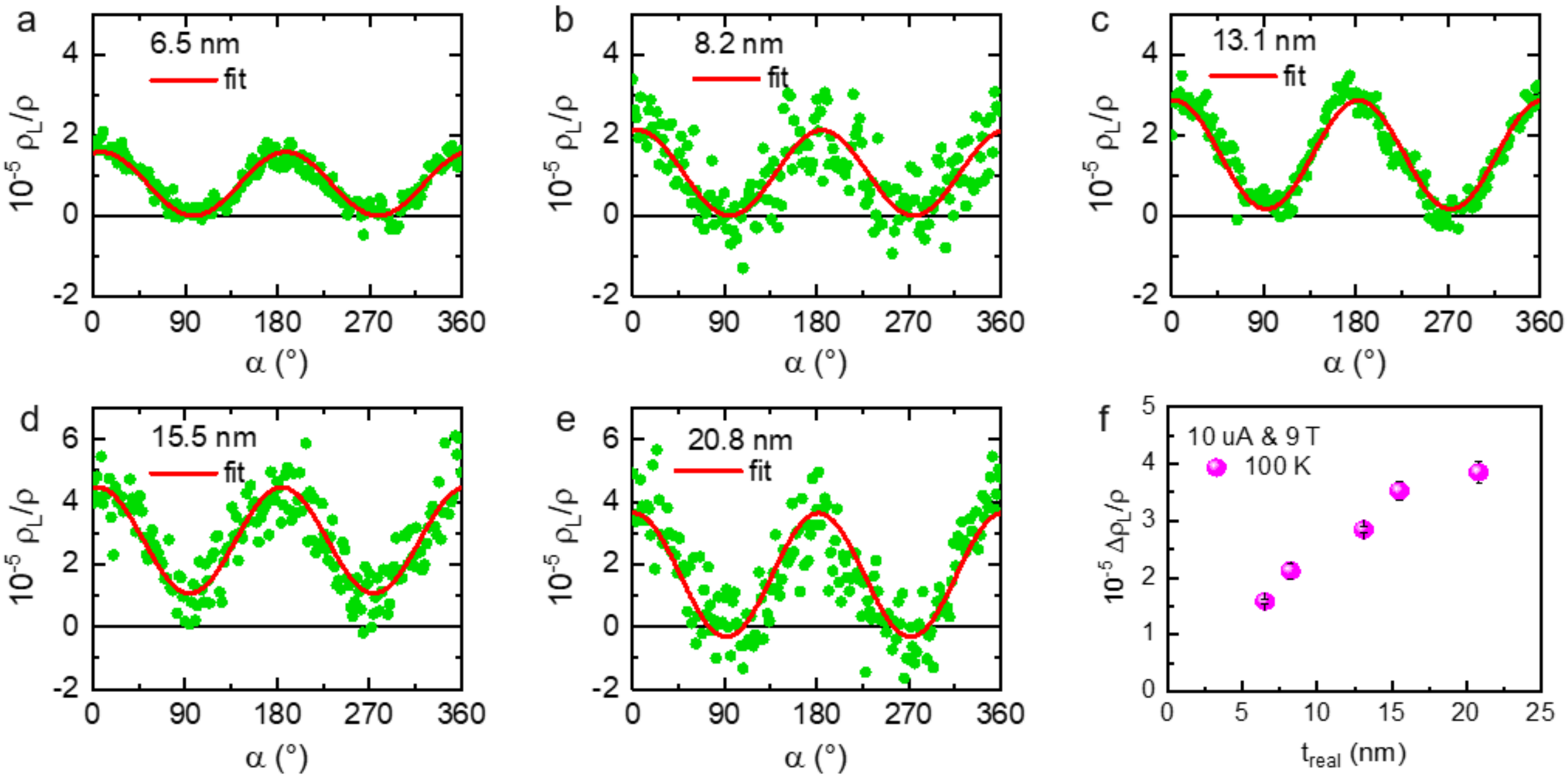

**Figure S5.** α-plane angle-dependent magnetoresistance of the SVO film series at T = 100 K and B = 9 T. The panels correspond to film thicknesses of (a) 6.5 nm, (b) 8.2 nm, (c) 13.1 nm, (d) 15.5 nm and (e) 20.8 nm; the 8.2 nm value is the interpolated thickness of the nominal 10 nm sample, as detailed in Table S1. The normalized longitudinal resistivity change is plotted as a function of the in-plane rotation angle α, and solid lines are $\cos^2$ fits. (f) Magnetoresistance amplitude $|A_\alpha|$ as a function of film thickness. All samples exhibit the same periodicity and phase, while the amplitude varies across the series.

## 6. Field-dependent magnetoresistance across the SVO thickness series

Field-dependent longitudinal magnetoresistance was measured with B ∥ x and B ∥ y for all investigated thicknesses, and with B ∥ z for selected films (Figure S6). The x-y comparison provides the common basis across the series: B ∥ x is transverse to L ∥ y, whereas B ∥ y is parallel to L. The z-oriented traces are included where available but are not used for the systematic thickness comparison.

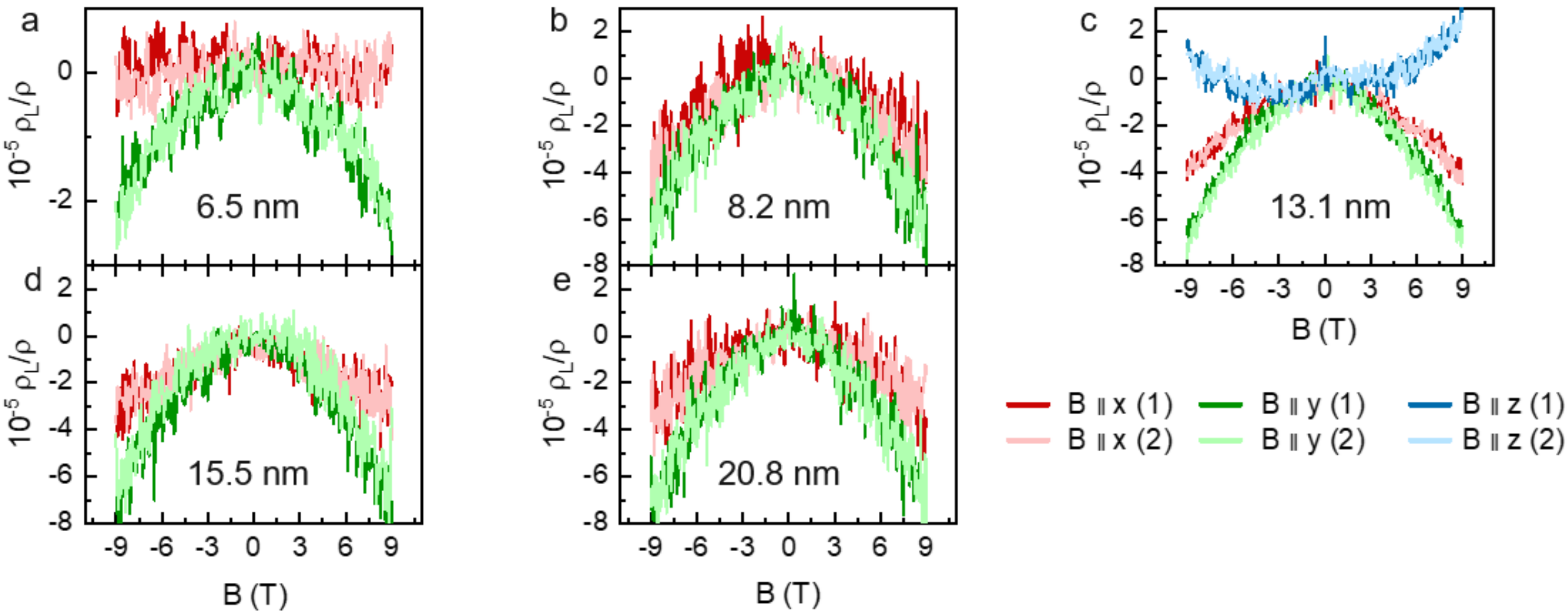

**Figure S6.** Field-dependent longitudinal magnetoresistance of the SVO film series at T = 100 K. B ∥ x and B ∥ y are shown for all investigated film thicknesses; B ∥ z traces are included where available. B ∥ x is transverse to L ∥ y, whereas B ∥ y is parallel to L. The x-y separation yields the positive, even-in-field HMR-like response discussed in the main text. The more sample-dependent B ∥ z response, including a positive high-field contribution where observed, is compatible with differences in crystalline order, impurity scattering, or defect landscape.

For every film, the x and y traces separate progressively with increasing |B|, and ρx(B) − ρy(B) is even in field, reproducing the HMR-like field evolution discussed in the main text. The B ∥ z response is more sample dependent and, where measured, can develop a positive high-field contribution. This variability is compatible with differences in crystalline order, impurity scattering, or defect landscape and is therefore treated as an additional background. The robust x-y response provides the field-sweep counterpart to the angular measurements in Figures S4 and S5.

## 7. First-principles computational details

[Final computational methodology by D.G. if necessary]

## 8. HMR model, orbital relaxation time, and sensitivity of the extracted orbital parameters

The quantitative analysis follows the diffusive HMR framework used in the main text. Within this framework, the field-dependent normalized longitudinal resistivity change is written as

$$\Delta\rho_L(B)/\rho = 2\theta_{OH}^2\{(\lambda_{OD}/t)\tanh[t/(2\lambda_{OD})] - \mathrm{Re}[(\Lambda/t)\tanh(t/2\Lambda)]\}, \quad \text{(S1)}$$

$$\Lambda^{-2} = \lambda_{OD}^{-2} + i\, g\, \mu_B B/(D_O \hbar), \quad \text{(S2)}$$

where t is the film thickness, $\lambda_{OD}$ is the orbital diffusion length, $D_O$ is the orbital diffusion coefficient, g is the Landé factor, μB is the Bohr magneton and Λ is the complex field-dependent diffusion length. The orbital Hall angle and orbital Hall conductivity are related through $\theta_{OH} = \sigma_{OH}\rho$. Following Ref. 12 in the main text, this relation defines $\sigma_{OH}$ in (ħ/2e) $\Omega^{-1}cm^{-1}$; values quoted below in (ħ/e) $\Omega^{-1}cm^{-1}$ are smaller by a factor of two.

The orbital relaxation time is related to the orbital diffusion coefficient by $\tau_{OD} = \lambda_{OD}^2/D_O$. Substituting $D_O = \lambda_{OD}^2/\tau_{OD}$ into Eq. (S2) separates the field-dependent precession scale from the diffusion length and gives

$$\Lambda^{-2} = \lambda_{OD}^{-2}[1 + i\omega_L\tau_{OD}], \quad \text{(S3)}$$

$$\omega_L = g\mu_B B/\hbar, \quad \text{(S4)}$$

where $\omega_L$ is the Larmor angular frequency associated with the applied magnetic field B. Once t, $\lambda_{OD}$, and g are fixed, the shape of the HMR field dependence is governed by $\tau_{OD}$, whereas $\theta_{OH}^2$ enters Eq. (S1) as an overall multiplicative factor. This makes it possible to constrain $\tau_{OD}$ from the relative field evolution without prior knowledge of $\theta_{OH}$.

For clarity, we define the field-dependent factor

$$F(B,\tau_{OD}) = (\lambda_{OD}/t)\tanh[t/(2\lambda_{OD})] - \mathrm{Re}\{(\Lambda/t)\tanh[t/(2\Lambda)]\}, \quad (S5)$$

so that $A(B) = 2\theta_{OH}^2 F(B,\tau_{OD})$. Taking the ratio of the α-plane amplitudes measured at 3 and 9 T cancels the unknown $\theta_{OH}^2$ prefactor:

$$R_{3/9} \equiv A_\alpha(3\text{ T})/A_\alpha(9\text{ T}) = F(3\text{ T},\tau_{OD})/F(9\text{ T},\tau_{OD}). \quad (S6)$$

For the representative 20.8 nm SVO film, the $\cos^2$ fits give $A_\alpha(3\text{ T}) = (4.68 \pm 0.68) \times 10^{-6}$ and $A_\alpha(9\text{ T}) = (3.96 \pm 0.19) \times 10^{-5}$. Direct propagation of the fit uncertainties gives $R_{3/9} = 0.1182 \pm 0.0181$. Numerical evaluation of Eq. (S6) with t = 20.8 nm, $\lambda_{OD}$ = 2 nm, and g = 2 gives the best estimate $\tau_{OD}$ = 0.201 ps, reported as 0.20 ps in the main text.

The asymmetric constraint on $\tau_{OD}$ follows from the quadratic low-field limit. For $\omega_L\tau_{OD} \ll 1$, expansion of Eq. (S1) gives, for fixed t and $\lambda_{OD}$,

$$A_\alpha(B) \propto \theta_{OH}^2\tau_{OD}^2B^2. \quad (S7)$$

Consequently, the amplitude ratio approaches a parameter-independent value as the response becomes purely quadratic:

$$\lim\,(\tau_{OD} \to 0)\; R_{3/9} = (3/9)^2 = 1/9 = 0.1111. \quad (S8)$$

The measured central ratio, 0.1182, is slightly above this quadratic-limit value and maps to $\tau_{OD}$ = 0.201 ps. The upper edge of the propagated fit-uncertainty interval, $R_{3/9}$ = 0.1363, maps to $\tau_{OD}$ = 0.387 ps ≈ 0.39 ps. The lower edge, $R_{3/9}$ = 0.1001, lies below the minimum ratio 1/9 allowed by the positive-$\tau_{OD}$ model. Therefore the present field-shape data do not resolve a finite lower bound on $\tau_{OD}$. The value $\tau_{OD}$ = 0.20 ps should be interpreted as the best estimate, with an upper compatible range of approximately 0.39 ps.

Importantly, the $\tau_{OD} \to 0$ statement in Eq. (S8) refers to the limiting shape of the normalized field ratio, not to a physically measured zero relaxation time. For finite $\theta_{OH}$, the absolute HMR amplitude simultaneously vanishes as $\tau_{OD}^2$ in this limit. Thus the finite measured curvature excludes an exactly zero HMR response, but it does not by itself establish a finite lower bound on $\tau_{OD}$ because the absolute low-field curvature depends on both $\theta_{OH}$ and $\tau_{OD}$.

The predominantly quadratic field dependence can be quantified independently from the field sweep. Fitting the 20.8 nm $\Delta\rho_{x-y}/\rho$ curve to

$$\Delta\rho_{x-y}/\rho = CB^2 \quad (S9)$$

gives $C = (5.93 \pm 0.59) \times 10^{-15}\ \mathrm{Oe}^{-2}$. In the low-field expansion of the HMR model, the quadratic coefficient has the general dependence

$$C \propto \theta_{OH}^2\tau_{OD}^2(g\mu_B/\hbar)^2 \times G(t/\lambda_{OD}), \quad (S10)$$

where $G(t/\lambda_{OD})$ is a dimensionless thickness-dependent factor set by the diffusion solution. Therefore a finite C establishes a finite quadratic HMR curvature but does not independently determine $\tau_{OD}$ unless $\theta_{OH}$ is known separately. The role of C here is instead to provide an independent consistency check that the measured response remains predominantly in the quadratic HMR regime, while $\tau_{OD}$ is constrained from the field-shape ratio in Eq. (S6). For reference, the fitted parabola gives $CB^2 = 5.34 \times 10^{-6}$ at 3 T and $4.80 \times 10^{-5}$ at 9 T, of the same order as the measured α-plane amplitudes, with the increasing deviation at 9 T consistent with the onset of departure from the strict low-field limit.

For the extraction of $\theta_{OH}$ and $\sigma_{OH}$, the saturation value is used only to establish a conservative lower bound. As B → ∞, the field-dependent second term in Eq. (S1) tends to zero, so the HMR amplitude approaches its maximum value

$$A_{sat} = 2\theta_{OH}^2(\lambda_{OD}/t)\tanh[t/(2\lambda_{OD})]. \quad (S11)$$

Because $A_\alpha(9\text{ T}) \le A_{sat}$, substituting the measured finite-field amplitude into the saturation expression gives a lower bound on the orbital Hall angle:

$$\theta_{OH} \geq \theta_{OH,LB} = \{A_\alpha(9\ T)/[2(\lambda_{OD}/t)\tanh(t/2\lambda_{OD})]\}^{1/2}. \quad (S12)$$

For $A_\alpha(9\ T) = 3.96 \times 10^{-5}$, $t = 20.8$ nm, and $\lambda_{OD} = 2$ nm, Eq. (S12) gives $\theta_{OH,LB} = 0.01435$. With $\rho(100\ K) = 15.6$ μΩ·cm and the conductivity convention stated above, this corresponds to $\sigma_{OH,LB} = 4.60 \times 10^2$ (ħ/e) $\Omega^{-1}cm^{-1}$.

The dependence of the lower-bound estimates on $\lambda_{OD}$ is evaluated explicitly rather than treating the reference value $\lambda_{OD} = 2$ nm as an independently measured quantity. For the 20.8 nm representative film, $A_\alpha = 3.96 \times 10^{-5}$ and $\rho(100\ K) = 15.6$ μΩ·cm give $\theta_{OH,LB} \approx 0.014$ and $\sigma_{OH,LB} \approx 4.6 \times 10^2$ (ħ/e) $\Omega^{-1}cm^{-1}$ when $\lambda_{OD} = 2$ nm. Varying $\lambda_{OD}$ from 0.5 to 5 nm changes the absolute lower-bound value of $\sigma_{OH}$ but leaves it within the same order of magnitude as the first-principles result. Table S2 summarizes the corresponding fixed-$\lambda_{OD}$ lower-bound estimates for the full series.

Table S2. Illustrative saturation-limit lower-bound orbital Hall parameters obtained from the α-plane amplitudes using the same reference orbital diffusion length $\lambda_{OD} = 2$ nm for all samples. The table is not used to extract a thickness dependence of $\lambda_{OD}$, because the longitudinal resistivity varies substantially across the series. $\theta_{OH,LB}$ is obtained from Eq. (S12). $\sigma_{OH}$ is reported in units of (ħ/e) $\Omega^{-1}cm^{-1}$ for direct comparison with the first-principles value. The thickness values follow Table S1; t = 8.2 nm* is an interpolated estimate rather than an XRR measurement.

| **Nominal t (nm)** | **t (nm)** | **ρ(100 K) (μΩ·cm)** | **$\|A_\alpha\|$ ($10^{-5}$)** | **F(t, $\lambda_{OD}$=2 nm)** | **$\theta_{OH}$** | **$\sigma_{OH}$ (ħ/e $\Omega^{-1}cm^{-1}$)** |
|---|---|---|---|---|---|---|
| 5 | 6.5 | 50.0 | 1.59 | 0.5694 | 0.00528 | 53 |
| 10 | 8.2* | 18.7 | 2.12 | 0.4719 | 0.00670 | 179 |
| 15 | 13.1 | 30.8 | 2.70 | 0.3045 | 0.00942 | 153 |
| 20 | 15.5 | 29.6 | 3.40 | 0.2578 | 0.01148 | 194 |
| 25 | 20.8 | 15.6 | 3.96 | 0.1923 | 0.01435 | 460 |

Table S2 should be interpreted only as a fixed-$\lambda_{OD}$ consistency exercise. It shows how the measured angular amplitudes map onto the saturation-limit lower-bound estimates $\theta_{OH,LB}$ and $\sigma_{OH,LB}$ when the same reference $\lambda_{OD} = 2$ nm is imposed for every sample. Because the samples differ strongly in resistivity, these values are not used to infer an intrinsic thickness dependence of the orbital Hall conductivity or of $\lambda_{OD}$. Table S3 summarizes the $\lambda_{OD}$ sensitivity for the 20.8 nm representative film.

Table S3. Sensitivity of the saturation-limit lower-bound orbital Hall parameters of the 20.8 nm representative film to the assumed orbital diffusion length. The values are calculated from Eq. (S12) using $A_\alpha = 3.96 \times 10^{-5}$, $t = 20.8$ nm and $\rho(100\ K) = 15.6$ μΩ·cm.

| **$\lambda_{OD}$ (nm)** | **$\theta_{OH}$** | **$\sigma_{OH}$ (ħ/e $\Omega^{-1}cm^{-1}$)** | **$\sigma_{OH}/\sigma_{OH,DFT}$** |
|---|---|---|---|
| 0.5 | 0.02870 | 920 | 2.36 |
| 1 | 0.02029 | 650 | 1.67 |
| 2 | 0.01435 | 460 | 1.18 |
| 3 | 0.01173 | 376 | 0.96 |
| 5 | 0.00922 | 295 | 0.76 |